# *In Situ* characterization of the proton enhanced conductivity of 20nm TiO2 thin-films obtained on the surface of optical fiber

Jacob L. Poole*, Paul R. Ohodnicki, Yang Yu, Jian Liu, Bret Howard, and Sittichai Natesakhawat

**Affiliations**

Dr. J. L. Poole, Dr. P. R Ohodnicki, Dr. Y. Yu, Dr. J. Liu, Dr. B. Howard, and Dr. S. Natesakhawat
National Energy Technology Laboratory, 626 Cochrans Mill Rd, Pittsburgh, PA 15236, USA
E-mail: JakePooleFin@gmail.com
Dr. Y. Yu
AECOM, 626 Cochrans Mill Rd, Pittsburgh, PA 15236, USA
Dr. S. Natesakhawat
Department of Chemical and Petroleum Engineering, University of Pittsburgh, Pittsburgh, PA 15261, USA

**Abstract**

We report *in situ* characterized $TiO_2$ thin-films deposited on optical fiber, having thicknesses in the 20-100nm range, and having enhanced conductivity values of 700S/cm upon interacting with hydrogen. This conductivity was achieved in pure hydrogen at 800-900°C, having a measured activation energy of 0.26eV of the hopping type. Given the variability in the observed results, it is postulated that the highest conductivity achievable may be much greater than what is currently demonstrated. The conductivity is retained after cooling to ambient temperatures as confirmed by Hall measurements, and subsequent grazing-incidence x-ray diffraction and TEM measurements show the films to be in the rutile phase. The exceptional conductivity in these films is hypothesized to result from direct proton incorporation into the lattice populating the conduction band with excess electrons, or from altering the Titania lattice to form conductive Magneli phases. The films did not display any evidence of transformations, however formation of Magneli phases was confirmed for powders. These interesting results, observed by examining 20nm films on the surface of optical fiber in combination with the first impedance spectroscopy performed on films on optical fiber in high temperature Fuel Cell type environments, confirm hypotheses arrived at in prior publications where thin-films of Titania had optical properties which could only be explained by the current claim. Titania thin-films on optical-fiber are being explored for high temperature hydrogen derived energy generation, thermo-photonic energy conversion, and associated sensors due to their unique interactions with hydrogen.

Energy conversion devices operating at high temperatures have numerous benefits such as higher efficiencies, water gas-shift reactions, fuel reforming, multi-fuel compatibility, and various thermally driven catalytic properties eliminating additional needed steps to accomplish these otherwise costly and critical processes. Furthermore, issues with compatibility, thermal stability, in addition to high costs pose stringent demands on useful material systems in the 700-900°C temperature range. Thus, materials with properties that improve upon the performance of high temperature energy conversion devices and sensors for the these systems are highly sought after, especially if the material is without substantial cost, complexity, and fabrication challenges.[1–4]

Hydrogen brings forth interesting material properties in films of Titania when coated on the surface of optical fiber, properties that may not be well understood. Hydrogen interactions can cause the formation of intermediate bandgap states due to mobile interstitial $Ti^{3+}$ and $H^+$ (**Figure 1**), making this simple oxide a highly functional material for numerous applications in photo-catalysis, photochemical water splitting, photo induced hydrophilicity, chemical sensors, and dye-sensitized solar cells.[5–13] N-type doping, by the introduction of donor species such as a pentavalent Nb to replace a trivalent Ti, also provide $Ti^{3+}$ useful in transparent conducting material exploration.[5,14,15]

Furthermore, recent demonstrations have shown considerable improvements in the extracted power density of Nafion based Fuel-Cells due to incorporating nanoparticles of $TiO_2$ into the electrolyte.[16–19] For which, the improvements were attributed to the hydration promotion by the nanoparticles, a claim supported by recent theoretical work.[20] Titania having an increased conductivity upon exposure to hydrogen has been suggested by work at lower temperatures and by theoretical studies.[20–22] In addition, recent optical measurements performed on $TiO_2$ films at 800°C in proton rich environments, for sensors and for thermo-photonic energy harvesting, has indirectly demonstrated that $TiO_2$ should have substantial electronic conductivity.[9,13,23,24]

In general, there is not a consensus on whether interstitial Ti or oxygen vacancies are the prominent defect species that form upon exposing $TiO_2$ to hydrogen at higher temperatures, or whether the electronic states are localized or delocalized.[13,25] There is some consensus on the polaron hopping nature of electron transport in $TiO_2$, however a wide range of mobility values and effective masses were reported.[7] It is suggested that the formation of deeper trap states by interstitial Ti leads to polaronic defects located approximately 0.8eV below the fermi level with an activation energy in the range of ~ 0.1eV, having fractional lifetimes in a large number of states.[20,26,27] In addition, high conductivity values have been reported for single crystal Magneli phases of Titania with the largest being for $Ti_4O_7$.[28] Being comparable in magnitude to what is reported here, where the high conductivity was attributed to having a reduced Ti to Ti separation yielding a reduced hopping activation energy on the order of ~0.013eV for electrons.[28] This behavior is suggested to be associated with hydrogen exposures at higher temperatures, forming interstitials that deform the crystal structure through shear planes.[13,25,29,30] While, other reports suggest hydrogen doping producing $Ti^{3+}$ species by $H_2$ disassociation into a proton bound to a lattice oxygen and a thermally excitable electron in shallow traps in the bandgap, as being responsible for the observed electronic conductivity.[20,31,32] Interestingly, it was also reported that proton diffusion can take place helically along the open c-axis of infrared irradiated rutile with an activation energy of 0.2eV, assisted by the O-H wag motion similarly to the Grotthuss mechanism of proton conduction.[31,33]

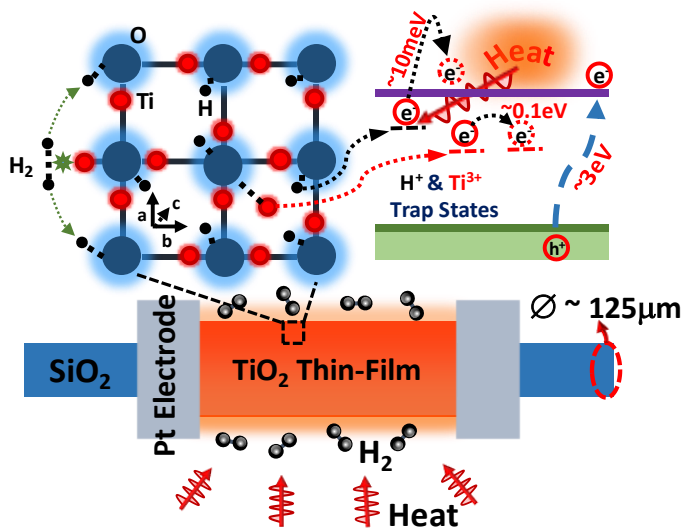

**Figure 1.** Schematic of a thin-film integrated silica optical-fiber, along with illustrations of some of the mechanisms thought to be responsible for the observed enhanced conductivity of proton-rich Titania. These are the formations of interstitial $H^+$ and/or $Ti^{3+}$, resulting in either an excess charge density in the conduction band and or charge hopping within the bandgap as a function of the environment thermal energy. Room temperature, being at 25meV, can easily break down the 10meV barrier for H+ trap states.

In this letter, we report measurements conducted using *in situ* Electrochemical Impedance Spectroscopy (EIS) on $TiO_2$ thin-films integrated with 105µm silica optical-fibers (**Figure 1**), along with Hall conductivity, TEM, and XRD measurements. These types of measurements, being done for the first time on films deposited on optical fiber, provide direct evidence and can be used for confirming hypotheses derived from other measurements, such as the previously reported optical measurements.

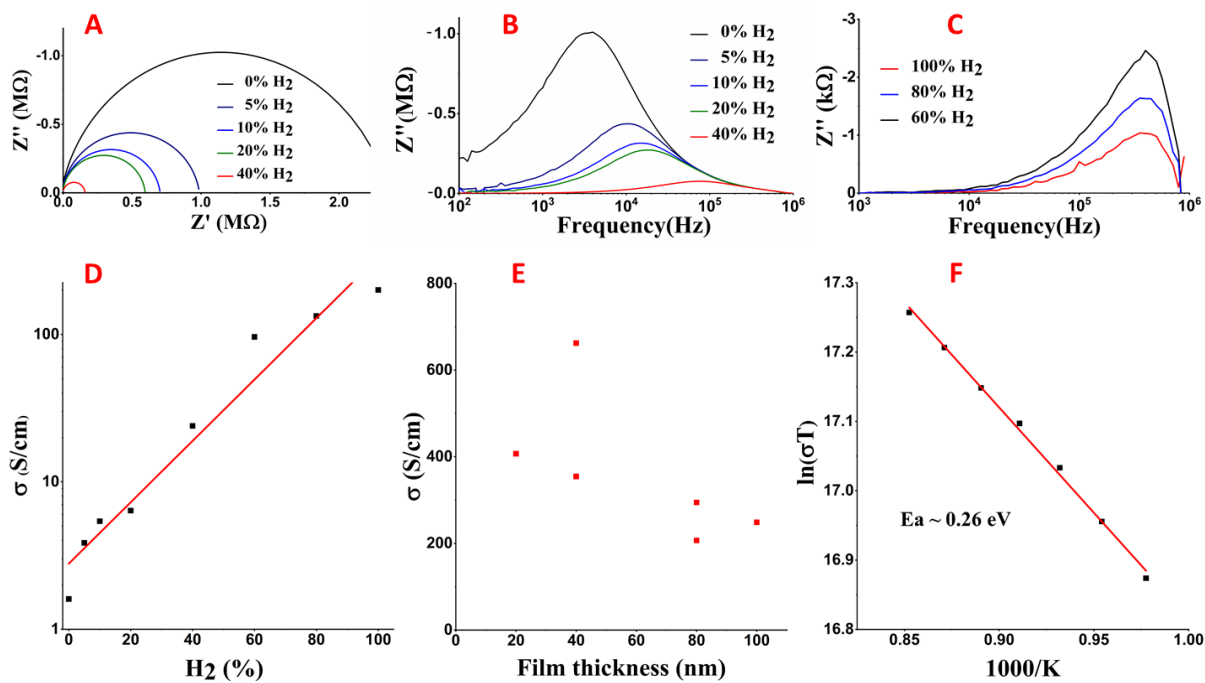

**Figure 2.** (**A**) Electrochemical impedance spectra of thin-films of $TiO_2$ on silica fiber measured at 900°C, showing proton induced conductivity enhancements of several orders of magnitude. (**B**)-(**C**) Bode plots showing the $H_2$ partial pressure dependent impedance. (**D**) DC conductivity as a function of $H_2$ partial pressure at 900°C. (**E**) Measured conductivity values of films of various thicknesses. (**F**) Arrhenius plot indicating a hopping-type activation energy ~0.26 eV.

The films were coated on the surface of optical fiber, as in the previous works, where the films were examined for high temperature thermal emissivity-based hydrogen sensors, and for high temperature hydro-thermo-photonic energy conversion.[9,23,24] While exploring the films on the surface of optical fiber for the discussed applications, indirect evidence for enhanced conductivity was observed, motivating interest in direct measurements of the electronic transport properties for comparison with the measured optical signatures. Combined measurements suggest that the interaction of $H_2$ with $TiO_2$ could be associated with direct $H_2$ uptake forming interstitial protons bound to lattice oxygens which serve as electron donor sites forming readily thermalized shallow electronic trap states with meV activation energy, imparting substantial electronic conductivity to this highly functional oxide even at room temperature, after quenching.[32] Other mechanisms are not expected to impart such a high electronic conductivity at room temperature due to the previously reported activation energy for hopping related conduction, which is on the order of ~0.1eV requiring thermal energy of ~850°C or greater to overcome this barrier. The high electronic conductivity attainable with this highly functional oxide is promising for applications spanning sensors, thermal energy harvesting, thermo-photonic energy conversion, fuel cells, and transparent conducting oxides, to name a few.

## Results and Discussions

Films of various thicknesses were coated onto the surface of 105μm diameter silica fibers and integrated with an alumina electrochemical probe with platinum contacts, designed to place the film inside of an environmental chamber at atmospheric pressure. The films were allowed to equilibrate for several hours under 100SCCM flow of hydrogen at a temperature of 900°C. The Electrochemical Impedance Spectrum (EIS) was measured with a Solartron in the 1kHz to 2MHz range, while the hydrogen partial pressure was adjusted from pure $H_2$ to an environment of only $N_2$, incrementally (**Figures 2A-C**). Conductivity enhancements of several orders of magnitude are clearly visible in the EIS spectra and the Bode plots due to successive increases of the $H_2$ partial pressure (**Figures 2A-C**). **Figure 2A** depicts the device's two component impedance plotted against one another with the frequency of excitation being a hidden dimension, whereas **Figures 2B-C** depict the capacitive impedance plotted versus the frequency of excitation of the film fiber composite. The frequency of electronic resonance displaces as a function of hydrogen partial pressure. The DC conductivity values (**Figure 2D**) were calculated from the real component of the EIS spectra (x-intercept of **Figure 2A** at 1 kHz) of the 40nm film, and is representative of the DC operating property of the film.

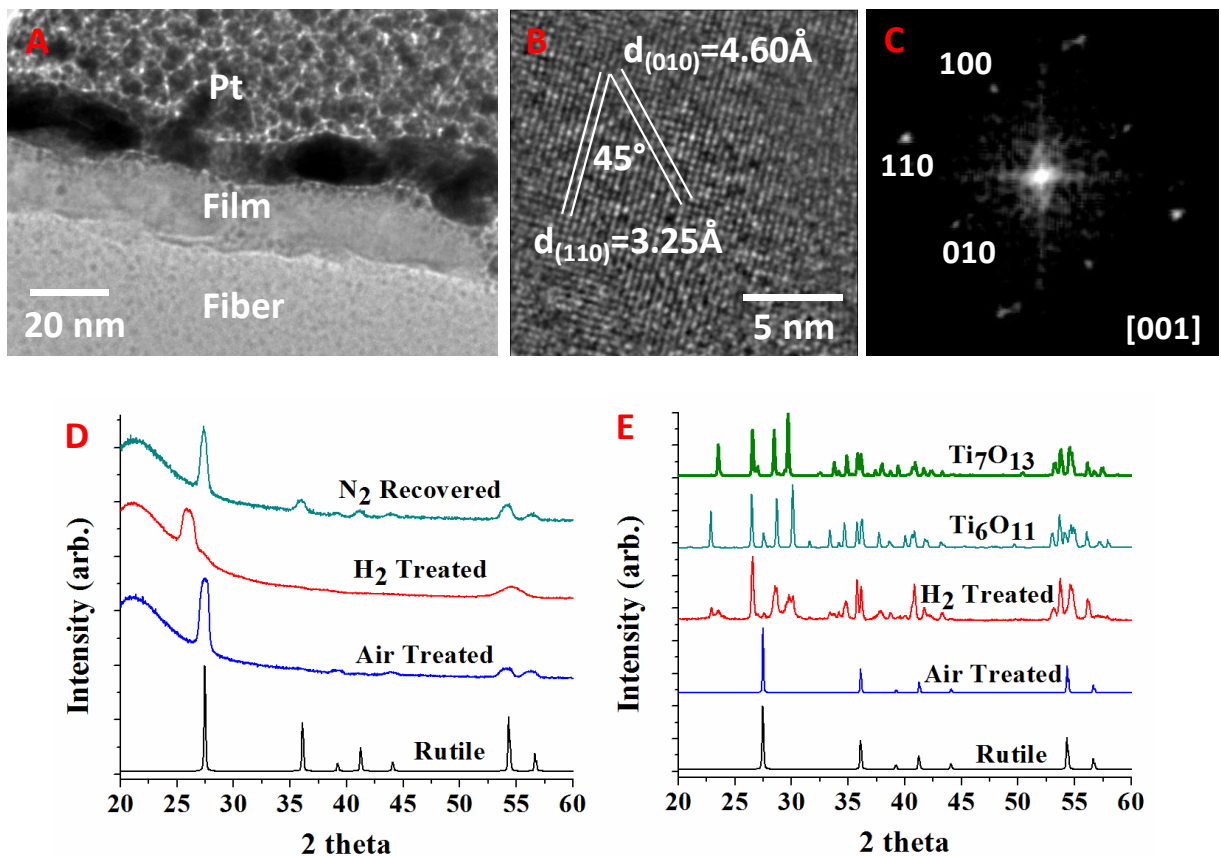

**Figure 3.** (**A**) Bright-field TEM image of the thin-film cross-section indicating a film thickness of ~20 nm for a single coating on a silica optical fiber. (**B**)-(**C**) High resolution TEM image with measured lattice spacing corresponding to the (110) and (010) planes with an FFT showing a rutile structure for the films coated directly on the optical fiber after an $H_2$ treatment at 900°C. (**D**) XRD spectra of thin-films on planar silica substrates annealed in air, then in $H_2$, and recovered in $N_2$ at 900°C, for which the $H_2$ annealed film shows a 2θ reduction of 0.144 in the lowest 2θ peak. (**E**) XRD spectra of powders annealed in $H_2$ and in air at 900°C. The powder spectra show the presence of Magneli phases to which $Ti_6O_{11}$ and $Ti_7O_{13}$ fit best.

A study of conductivity as a function of the film thickness resulting from H$_2$ exposure (**Figure 2E**), shows deviations of factors of 2 or 3, which is consistent with the expected variability of solution phase processing to form functional thin films. Nevertheless, the solution processing route allows for accommodating various geometries, as in this case the film was deposited onto the surface of 125µm diameter cylindrical silica optical fiber. Moreover, the optical fiber is a highly useful platform for sensor exploration and novel thermo-photonic energy harvesting methods.[9,23,24] The Arrhenius plot obtained in H$_2$ (**Figure 2F**) suggests an activation energy of ~0.26 eV of the hopping type for the temperature dependence of measured electronic conductivity, which is larger than expected for polaron hopping (~0.1eV) based on prior reports but it is the correct magnitude for proton diffusion within the TiO$_2$ network. This suggests a thermal energy driven proton incorporation, imparting electronic conductivity to the film. The environmental thermal energy promotes the disassociation of hydrogen both in the ambient and at the surface of TiO$_2$, followed by proton incorporation into the lattice. For example, an activation energy of 0.2eV for proton hopping was given in a previous report for an IR irradiated single crystal of TiO$_2$ along the c-axis, where it was reported that incorporated protons were highly mobile through wag assisted motion, along the open c-channel.[31,33] For a polycrystalline thin film, the microstructure is thought to be responsible for the increase in activation energy from 0.2 to 0.26eV, in comparison to what was reported for single crystals for the Grotthuss mechanism, which is likely due to having an increased O-O separation along other directions.

With platinum assisted Focused Ion Beam (FIB) a portion of a film was lifted directly from the optical fiber after a reduction treatment in pure hydrogen at a temperature of 900°C, demonstrating that a single film coating has a thickness of ~20 nm (**Figure 3A**). The high resolution TEM image shown in **Figure 3B** along with the FFT (**Figure 3C**) show evidence for a rutile TiO2 structure. The lattice spacings identified in **Figure 3B** correspond to the (110) plane (d-spacing ~ 3.25angstroms) and the (010) plane (d-spacing ~ 4.60angstroms) for a pattern that can be indexed as an [001]-type zone axis.

The XRD spectra prepared on planar silica substrates were observed to be relatively weak, as would be expected for 20nm films, and were obtained with some difficulty. For a sample in which the crystallinity was resolved through XRD after annealing at 900°C in air, the observed XRD pattern conforms to the expected rutile peaks with some degree of preferred orientation due to relatively large intensities of the (110) and (220) peaks. After annealing the film in pure H$_2$ at 900°C, a shift in the position of the lowest 2θ peak (2θ ~ 27.44) by 0.144 was observed, corresponding to an increase in the a-axis lattice parameter. Upon further treatment of the hydrogen annealed film in air or nitrogen at 900°C, the positions of the XRD peaks returned to the original positions, along with a corresponding reduction in the degree of preferred orientation as illustrated by the "N2 treated" XRD pattern in **Figure 3D**. Such observations are consistent with direct uptake of H$_2$ within the films, resulting in an increase in the effective lattice parameter of the TiO$_2$ rutile phase. It has been reported that an incorporated proton in TiO$_2$ Rutile is responsible for the creation of an electron populated shallow trap state in the conduction band with an activation energy of ~10meV, which should readily thermalize even at room temperature [30] In contrast, the XRD spectra taken on powders is shown in **Figure 3E**, alongside of the two Magneli phases that the hydrogen exposed TiO$_2$ powder has spectra closest to. The powder treated in the same manner first in air align well with the reported Rutile structure. However, powders treated in hydrogen exhibit peaks that align best with the two Magneli phases whose XRD peaks are shown in **Figure 3E**. This hypothesis is further substantiated with the experimentally measured temperature dependency of the electronic conductivity as an H$_2$ annealed film quenched to room temperature retains the conductivity enhancement. For a 200nm film deposited on a quartz substrate, this was measured to be σ ~ 294S/cm with a carrier concentration n of $3.6 \times 10^{20}$ cm$^{-3}$ and mobility µ of 5.1 cm$^2$V$^{-1}$s$^{-1}$, measured with a Hall probe. While, another sample was measured to have a conductivity value of σ ~ 700 S/cm. This value is in sharp contrast to an air annealed film having σ=$5.87 \times 10^{-4}$ S/cm, n=$1.24 \times 10^{16}$ cm$^{-3}$, and µ=0.3 cm$^2$V$^{-1}$s$^{-1}$. Significant variability was observed between different prepared films, consistent with the deposition method being simple and benchtop in ambient

In conclusion, substantial electronic conductivity can be imparted to TiO$_2$, on the order of 700S/cm, by heating in hydrogen at temperatures of 800-900°C, having a measured charged carrier hopping activation energy of 0.26eV alligning with the published results of ref. 31 and 33, associating this energy with hydrogen diffusion. Hall measurements indicate that upon quenching the films retain their imparted conductivity values, suggesting the formation of H$^+$ induced shallow electronic trap-states in the bandgap. The promotion of an electron to the conduction band from the trap-state has an energy below that of room temperature (25meV), since the conductivity is retained even at room temperature for an indefinite amount of time. TEM and XRD measurements show evidence for a rutile phase and a lattice expansion of protonated samples, consistent with an H$^+$ uptake driven mechanism for films, which along with the measured activation energy points to a Grotthuss mechanism type action. While, XRD performed on the powders show the formation of Magneli phases. Therefore, 20nm thin-films of TiO$_2$ are highly functional for hydrogen driven high temperature energy systems and sensors for those systems due to their ability to solubilize protons. A recent demonstration showed that these films are highly desirable for applications such as thermo-photonic energy conversion.[9,23,24]

**Experimental Section**
*Sample Preparation*: An alkoxide route was adapted for the TiO$_2$ thin-film preparation followed by coating onto silica fibers (Thorlabs FG105LCA), as previously reported.[9,34] A 4-bore alumina tube was used as the test fixture with four platinum wires as the leads. A 1-cm long thin-film coated fiber was mounted to contacts with platinum paint onto one end of the alumina tube, then placed inside a controlled atmosphere quartz tube within a tube furnace.

*Sample Characterization*: The film Electrochemical Impedance Spectra was measured with a Solatron SI 1260 and 1287, *in situ,* underflow of $H_2$ at various partial pressures, and in the frequency range of 1 kHz to 2 MHz. A Panalytical X'Pert Pro diffractometer was used to collect the diffraction patterns a $TiO_2$ films at a grazing incidence angle of 2°. The room temperature conductivity of the films, carrier concentration, and mobility were measured using an Ecopia HMS-7000 Hall system. FIB lift-out was conducted with an FEI Nova Nanolab 600 dual beam FIB with a gallium ion beam, after coating the surface with platinum to maintain film integrity. BF TEM and HRTEM were conducted with a 200kV FEI Tecnai F20 TEM.

**Acknowledgements**
This work was funded by the U.S. DOE Advanced Research/Crosscutting Technologies program at the National Energy Technology Laboratory. This research was supported in part by an appointment to the National Energy Technology Laboratory Research Participation Program, sponsored by the U.S. Department of Energy and administered by the Oak Ridge Institute for Science and Education.

This report was prepared as an account of work sponsored by an agency of the United States Government. Neither the United States Government nor any agency thereof, nor any of their employees, makes any warranty, express or implied, or assumes any legal liability or responsibility for the accuracy, completeness, or usefulness of any information, apparatus, product, or process disclosed, or represents that its use would not infringe privately owned rights. Reference herein to any specific commercial product, process, or service by trade name, trademark, manufacturer, or otherwise does not necessarily constitute or imply its endorsement, recommendation, or favoring by the United States Government or any agency thereof. The views and opinions of authors expressed herein do not necessarily state or reflect those of the United States Government or any agency thereof.

**Data Availability**
Raw data were generated at the National Energy Technology Laboratory Pittsburgh facility. Derived data supporting the findings of this study are available from the corresponding author upon reasonable request.